\documentclass[a4paper,floatfix,rmp,twocolumn,showkeys,superscriptaddress]{revtex4}
              \usepackage[latin1]{inputenc}
\usepackage{graphicx}

\begin{document}

\title{Synthetic circuit designs for Earth terraformation}

\providecommand{\ICREA}{ICREA-Complex Systems  Lab, Universitat Pompeu
  Fabra   (GRIB),   Dr    Aiguader   80,   08003   Barcelona,   Spain}
\providecommand{\IBE}{Institut de Biologia Evolutiva, CSIC-UPF, Pg Maritim de la Barceloneta 37, 08003 Barcelona, Spain}
\providecommand{\SFI}{Santa Fe  Institute, 1399 Hyde  Park Road, Santa
  Fe  NM   87501,  USA} 
  
\author{Ricard V. Sol\'e
\footnote{Corresponding author}}
\affiliation{\ICREA}
\affiliation{\IBE}
\affiliation{\SFI}   
\author{{Ra\'ul Monta\~{n}ez}}
\affiliation{\ICREA}     
\affiliation{\IBE}
\author{Salvador Duran-Nebreda}
\affiliation{\ICREA}
\affiliation{\IBE}         

\vspace{0.4 cm}
\begin{abstract}
\vspace{0.2 cm}  
{\bf Background} 
Mounting evidence indicates that our planet might experience runaway effects associated to rising temperatures 
and ecosystem overexploitation, leading to catastrophic shifts on short time scales. Remediation scenarios capable 
of counterbalancing these effects involve geoengineering, sustainable practices and carbon sequestration, among others. None of these scenarios seems powerful enough to achieve the desired restoration of safe boundaries.
\vspace{0.2 cm}  

{\bf Hypothesis} 
We hypothesise that synthetic organisms with the appropriate engineering design could be used to safely prevent 
declines in some stressed ecosystems and help improving carbon sequestration. Such schemes would include 
engineering mutualistic dependencies preventing undesired evolutionary processes. We hypothesise that some 
particular design principles introduce unescapable constraints to the engineered organisms that act as effective firewalls.

\vspace{0.2 cm}  

{\bf Implications} 
Testing this designed organisms can be achieved by using controlled bioreactor models and accurate computational 
models including different scales (from genetic constructs and metabolic pathways to population dynamics). Our 
hypothesis heads towards a future anthropogenic action that should effectively act as Terraforming agents. It also 
implies a major challenge in the existing biosafety policies, since we suggest release of modified organisms 
as potentially necessary strategy for success.

\end{abstract}

\keywords{Synthetic Biology, Ecological Engineering, climate change, catastrophic shifts, mutualism}

\maketitle

\begin{quote}
\begin{flushright}
{\em The future cannot be predicted,\\ but futures can be invented}\\
Dennis Gabor
\end{flushright}
\end{quote}

\section{Background}

Climate change, along with a rapid depletion of natural resources and biodiversity declines 
is driving the biosphere towards unstable states. Widespread evidence indicates that 
increasing rise of average temperatures is leading 
to local, regional and global modifications of extant habitats, seriously endangering 
the future of our planet [1,2]. Given the large scale of the problem, suggested scenarios 
based on human intervention might fail to properly address the ongoing changes. 
Additionally, the time evolution of these changes can rapidly accelerate due 
to runaway effects associated to the nonlinear nature of these phenomena. In other words, 
current continuous changes might end up in so called {\em catastrophic shifts} [3-5]. Are we 
going to be capable to avoid them? 

Along with a better understanding of these changes, scientists and engineers have also 
come up with potential remediation scenarios to ameliorate and even stop the current 
trends. Different strategies involving mitigation [6] geoengineering [7-9] or adaptation [10] have been 
proposed. Mitigation implies measures that slowdown ongoing emission rates or 
provide ways for limiting emissions while geoengineering explicitly requires directed change. Geoengineering 
has been questioned due to staggering costs, unknown outcomes and limited impact (particularly in relation with 
$CO_2$) which make unclear their potential for counterbalancing current trends [7,11,12]. Adaptation scenarios 
place us in a future world where we will need to cope with new environmental and economic constraints. 
None of these suggested solutions might be a definite solution, but clearly the price for inaction will be much 
larger than any of the previous possibilities. 

It has been recently suggested that an alternative possibility would involve actively 
acting on the biosphere through the use of synthetic biology [13]. This approach 
could be used, among other things, as a way to curtain the accumulation of 
greenhouse gases, enhance nitrogen fixation or slow down degradation 
in arid and semiarid ecosystems. The key point of this proposal is that engineering living systems allows 
to reach large scales thanks to the intrinsic growth of the synthetic organisms. This makes 
a big difference in relation to standard engineering schemes, where artefacts need 
to be fully constructed from scratch. Instead, once a designed population is released, 
appropriate conditions will allow the living machines to make copies of themselves and expand to the 
desired spatial and temporal scales.

This approach, which is an effective way of ``Terraforming" the biosphere, needs 
to consider potential scenarios that guarantee an efficient result as well as a limited 
evolutionary potential. Designed microbes capable of functioning only under 
specific conditions have been constructed and strategies to incorporate genetic safeguards 
explored [16]. One avenue, to be used in biomedical applications, is to force the 
need for xenobiotic (unnatural) molecules that need to be supplied along with the 
genetically modified bacteria [17]. In this context, target habitats for designed organisms should 
be chosen as an additional, ecological-level containment strategy. Moreover, 
limits to the impact of synthetic organisms can be obtained using ecological interactions that 
are based on either cooperative loops or habitat constraints that are specially well met 
by different classes of anthropogenic-modified scenarios. In this paper we consider four possible engineering motifs 
that can cope with these two constraints. We do not consider explicit case studies (i. e. 
detailed genetic constructs or designed organisms) but instead the logic design schemes.

\section{Presentation of the hypothesis}

The obvious criticism to the scenario presented in [15] has to do with the unknown consequences of 
ecological and evolutionary dynamics on the engineered ecosystems. 
Actually, it can be argued that well known cases of exotic species 
introduced in some ecosystems caused large-scale disasters [18,19]. The list includes 
the introduction of different kinds of species into a novel habitat where they 
have benefited from a higher efficiency to exploit available resources. 
This situation corresponds (at least transiently) to a population positive feedback loop 
that involves an accelerated expansion (typically exponential in its first phase). 
Is there a rational strategy that can minimise the impact of an engineered species?

One way of preventing undesired explosive growth is to use a modified 
version of an extant organism that exhibits a strict relationship with another species 
associated to the target habitat. This means engineering a strong ecological link 
that makes spread limited. That would result in 
population dynamical processes preventing undesired growth of the modified organism. 
Moreover, using the appropriate context, strong habitat constraints can act in synergy 
as ecological firewalls. 

Here we suggest that two main avenues can be followed. 
One is engineering mutualistic relationships with resident organisms through the modification of 
already extant microorganisms or fungi. Recent experimental studies indicate that 
such designed mutualistic link can be created by artificially forcing a strong metabolic dependence 
and also with the help of genetic engineering. These studies have shown that 
the end product can be a physically interacting, stable pairwise relationship.

\begin{figure*}
{\centering \includegraphics[width=13cm]{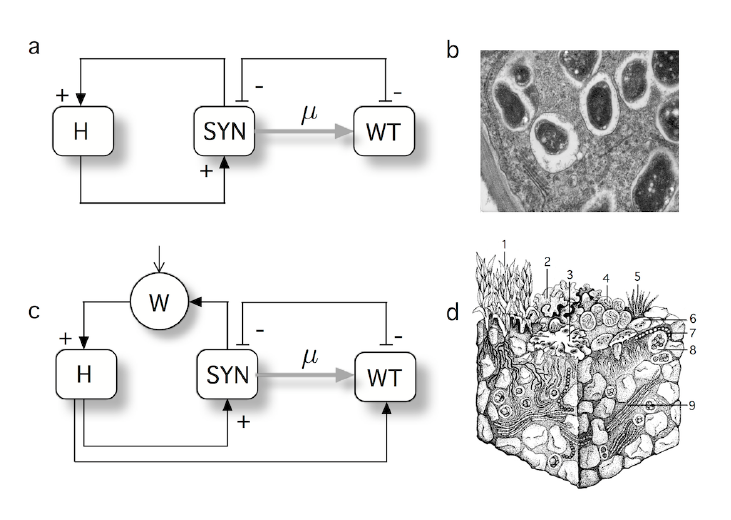}
\caption{
Terraformation motifs involving closed cooperation among players. Two main classes of 
     potential engineered synthetic microbes (SYN) interacting with their hosts (H) are indicated. Assuming that the engineered 
     species has been obtained from an existing one in the same environment, the wild type (here indicated as WT) 
     can be obtained from SYN if the engineered construct is lost by mutation (here indicated as a fray arrow, 
     and as  a rate $\mu$)
     In (a) we display a logic diagram of positive interactions among both partners defining a 
     mutual dependency. In (b) such cooperative interaction is mediated through some class of 
     physical factor, such as water (W). These two classes correspond, for example, to exclusive 
     mutualistic interactions displayed by plant cells within root nodules (c) where nitrogen-fixing bacteria 
     are physically embedded (image from wikipedia commons). 
     On the other hand, the need for survival under stressful conditions, 
     as those common in arid ecosystems, makes water a major player and limiting resource. 
     An engineered microbe capable of improving moisture retention can have a very strong effect 
     on the underlying plant species, expanding their populations. In soil crusts (d) a whole 
     range of species exist, adapted to water-poor conditions (drawing adapted from Belnap et al 2001). 
     Here we indicate (1) mosses (2,3) lichens, (4,5,7,9) cyanobacteria, (6) fungi and (8) green algae.}
\label{BB1}
}
\end{figure*}

Another possibility is to use a modified organism that grows on a given waste-related 
substrate that can be preferentially (or exclusively) used, and may be degraded, 
by the synthetic organism. Such substrate can 
be plastic garbage, sewage and other sources of human-created waste. 
Additionally, some special habitats might be ideal to grow 
strains of engineered microbes capable of performing a given functional task and unable to survive 
outside their restricted environment. 

In the next section, we consider a list of candidate engineering designs (and their 
variants) that could fit the 
previous description. We will define their basic logic and outline potential scenarios 
for their implementation, as well as potential drawbacks.

\section{Synthetic Terraformation motifs}

In this paper we introduce four potential bioengineering schemes. Hereafter, H and SYN indicate 
the target host and a synthetic microbe, respectively. Here SYN might have been obtained from some 
existing wild type strain (WT). Similarly, R is used to indicate some sort of 
resource, while W stands for water. The basic designs are intended to represent 
the logical organisation of our proposed constructs, and not the specific genetic 
designs. For this reason, since they are introduced as logic graphs, we choose 
to call them {\em Terraformation motifs} (TMS) to indicate this logic nature. 

The first two motifs deal with the engineering of 
cooperative interactions, either directly or indirectly. The third incorporates a design principle 
grounded in a tight dependence of the engineered microbes with a specific class of available 
resource or physical support. The fourth involves the use of an existing, human-generated 
waste habitats as the substrate of engineered microbes, will be controlled through some 
class of lethality outside their niche.

\subsection{Engineered mutualism} 

In this case, an engineered candidate organism is used to modify 
ecological systems through an engineered mutualistic relation in such a way that 
it will spread only if associated to its mutualistic partner. Mutualism requires a double positive feedback where 
the synthetic species benefits -and is benefited by- its host. Ideally,  failure should end in the 
disappearance of the modified species. In figure 1a we display the TM associated to this 
approach. Here the host and the synthetic organism have been designed to enhance each 
other's growth. Moreover, the synthetic species has been derived from an existing wild type 
strain and it can thus mutate into WT. This will be the case if the engineered part is not 
enough advantageous and instead becomes a burden for the microorganism. 

This scenario is tied to the symbiotic relationships that characterise several types of 
natural associations, such as nitrogen-fixing bacteria living in plant root nodules (fig 1b). 
Several experimental approaches have shown that such mutualistic relationship can be enforced 
by co-evolving plants and bacteria under strong selection together with 
genetic engineering. Engineering mutualistic symbiosis is already a reality. Proper 
manipulation of free-living species allow to force them to become obligate mutualists. 
This includes synthetic cooperative strains [20] evolving a plant pathogen into a legume symbiont [21,22], fungal-plant 
mycorrhizal symbiosis [23] yeast-alga and fungi-alga associations created through a forced environmental change [24] or 
by means of long-term selection experiments enforcing metabolic dependencies [25] among others.

\subsection{Indirect cooperation} 

Cooperation can also arise from an interfaced interaction where 
one of the species modifies the existing medium in such a way that the partner can thrive 
and create more growth opportunities for the first. The canonical example can be a species 
of microbe that has been engineered to excrete a molecule capable of enhancing water retention 
in arid conditions (Figure 1c). Here a microbe that exists in the chosen context can be 
engineered in order to release some kind of protein capable of enhancing water retention. Potential 
candidates would be engineered cyanobacteria that are known to produce 
extracellular polysaccharides [27,28]. Enhanced production of these molecules 
by synthetic strains could easily improve dry land soils and yield. The soil 
crust in particular (figure 1d) involves a rich ecosystem composed by lichens, 
mosses and cyanobacteria [29, 30] and constitute a crucial regulator of soil respiration in dryland 
ecosystems. Strategies oriented to soil rehabilitation and carbon sequestration could be implemented through 
 the engineering of soil crust  [31,32].

\begin{figure}
{\centering
  \includegraphics[width=7.3 cm]{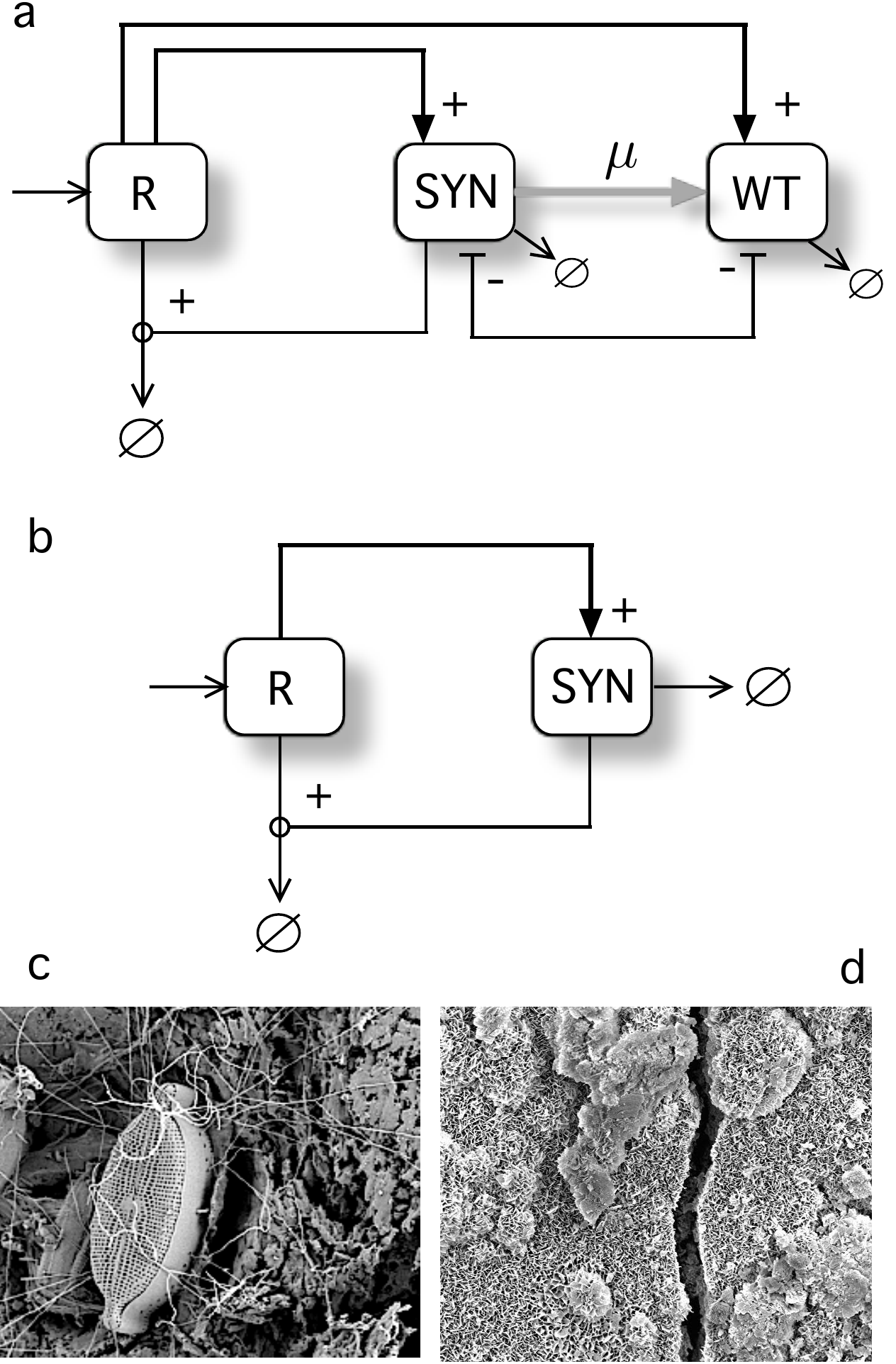}
\caption{Function-and-die Terraformation motif. Here a given substrate $R$ is being generated 
at a given rate and provides physical substrate to the synthetic population. The TM motif in (a) is 
based on the modification of an extant species, whereas in (b) we just assume that the 
engineered species has been improved to attach efficiently to the substrate. in both cases, 
the engineered species could perform a function while degrading the waste material. Candidate examples 
are plastic ocean debris, where many species are known to live (c) or concrete cracks (d). Figures (c) and (d) have been 
adapted from [43] and [44], respectively.}
\label{BB2}}
\end{figure}

In an arid ecosystem, plants can improve their growth thus expanding their population 
and providing further opportunities for microbial populations also to grow. In arid and semiarid 
habitats, plants typically develop local interactions involving so called {\em facilitation}: the presence of 
neighbouring plants favours the establishment of others and the preservation of 
a healthy soil [33]. Given the constraints imposed by water shortage and overgrazing, patchy distributions of 
plants are the common pattern [34-37]. Mounting evidence suggests that the conditions allowing these 
ecosystems to survive and the nonlinear nature of facilitation implies the existence of 
breakpoints and catastrophes: once reduced water availability or grazing pressure cross a given threshold, a rapid transition to the desert state should be expected. Modified organisms capable of building the indirect co-operative loop outlined above would easily increase facilitation. The increasing role of arid and semi-arid ecosystems 
as carbon sinks [38] makes them a specially relevant target for our terraformation proposal.

\begin{figure}
{\centering
  \includegraphics[width=8 cm]{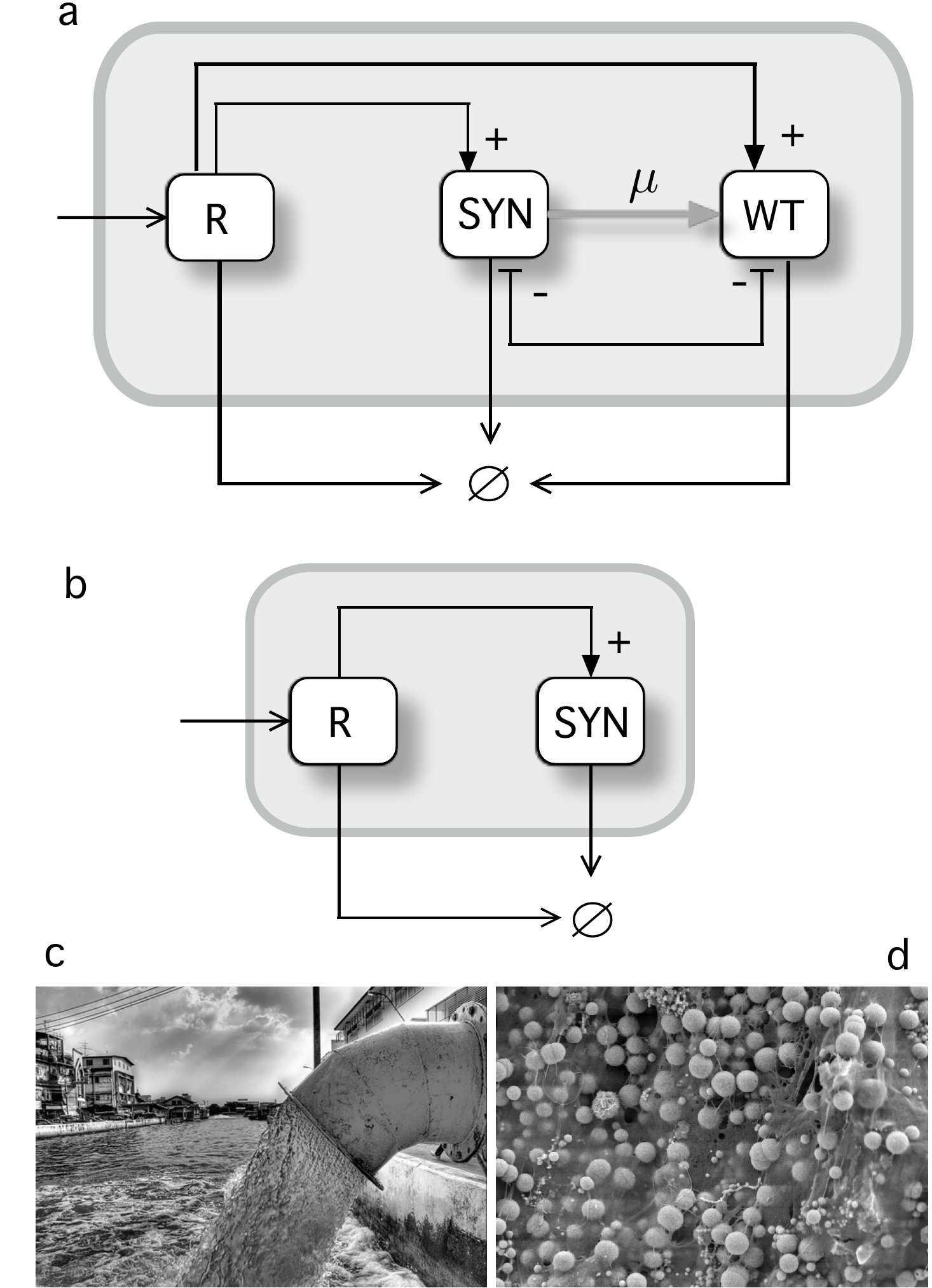}
\caption{Sewage-based terraformation motif. In (a) we consider a situation where 
an artificial environment is created as a byproduct of human activities producing waste. 
Our two strains are both sustained by available nutrients and physical conditions but now 
all of them are removed (burned or released) at a given rate. A simpler alternative (b) does 
not require engineering of extant species. A typical scenario would be sewage-related infrastructures (c) 
where a rich microbial community (d) is known to exist.}
\label{BB3}
}
\end{figure}

\subsection{``Function and die" design}

An engineered microbe performing a given functionality (such as carbon sequestration) 
can be coupled to the degradation of a given resource, such as plastic garbage or other long-living 
byproducts of human activities. This scenario is strongly tied to 
the problem of bioremediation [39-40] Here a non-living resource ($R$) is produced 
from anthropogenic actions and it provides the physical substrate where individuals can 
attach. In figure 2a we consider a TM that follows our previous scheme (again, a synthetic strain is derived form 
an existing one). In this case, however, no mutualistic loop is at work. Instead, both SYN and WT 
would attach to the substrate $R$ and thus their populations deepen on such potential for adhesion, 
which could be improved in the designed strain. 

A good candidate could be plastic garbage in the ocean [41,42] which is known to be 
colonised by many different species, including several microbial genus, such as Vibrio [43]. In this context, 
it is worth noting that, despite the rapid increase in plastic waste dumped in the ocean, the observed 
amount of plastic in open waters is much less than expected  [44] suggesting (among other possibilities) 
that some microbial species capable to attach to plastic polymers are also degrading them. 
This observation indicates that evolutionary forces might have favoured plastic-eating strains 
which could be used as engineering targets. If the only goal of the SYN is degrade 
the waste material, we could use a modified organism 
that might not be normally attached to this substrate (figure 2b). Different species, both prokaryotes 
and eukaryotes, are known to persist in plastic (figure 2c). Since removal of plastic debris might actually 
part of the goal, it might be unnecessary to use existing species associated to this substrate. Instead, 
it could be more efficient to simply design or evolve a highly-efficient species capable of attaching 
to the plastic surface and to over-compete other species.  

A different scenario that can be represented by our motif is provided by engineered bacteria that can be used to repair 
concrete cracks (figure 2d). The alkaline environment makes difficult for most species to thrive but 
some species can be used to this purpose [45]. Here the designed bacteria would enter, grow and replenish 
cracks with calcium carbonate until the task is finished. Several strategies have been used to this end and 
major improvements have been obtained [46,47]. A major advantage of this problem is that anaerobic 
bacteria are not going to survive outside the crack and thus selection immediately acts once the task is finished. Once again, 
the right combination of genetic design and ecological constraints create a powerful safeguard against 
undesired evolution.

\subsection{Sewage synthetic microbiome}

Urban centres are the largest human structures and as such 
they also incorporate massive infrastructures associated to treatment of waste as an end part of the 
city metabolism. Sewage systems and landfills offer a specially interesting opportunity to apply our approach. 
It is known that sewage systems involve their own micro biomes [48] and that some evolved microbes 
are currently causing damage to the concrete [49]. On the other hand, the sewage-based scenario is specially 
useful in our context, since microbes are eventually removed once they reach the open sea due to 
osmotic shock. If the same basic scheme is used, namely engineering an existing species, the TM  
can be summarised in figure 3a. Here a constant removal of both waste waters and microbes is represented 
by the arrows ending as $\rightarrow \emptyset$. 

Here too it might be less relevant to preserve the existing 
species of microbes, thus making unnecessary to engineer from wild type (figure 3b). Being part of the 
human infrastructures of developed countries (figure 3c) the sewage TM is also relevant to asses the 
potential dynamical responses of bioengineered ecosystems. The existing sewage and urban 
microbiomes (figure 3d) provide a rich repertoire of candidate species, although we just start to 
grasp their richness [50]. An interesting connection between these potential engineered strains and the 
gut microbiome has been pointed in [13]. The later defines an enormously rich microbial ecosystem that 
has coevolved with our species through our long evolutionary history. Ongoing biomedical research starts 
to be oriented towards intervening in the microbiome by means of both drugs but also microbial strains 
that might act like exotic invaders aimed to restore lost functionalities [51-53].

\section{Discussion}

The three major classes of TMs presented above provide a framework to design synthetic 
biology alternatives to existing strategies aimed to fight against climate change 
and its consequences. A main departure from geo-engineering is the fact that designed 
living machines are by definition capable of self-replication. From an engineering perspective, that implies that 
the designed biomachines will be capable of making new replicas and thus scale up the problem. The 
synthetic organisms associated to the TMs act as {\em ecosystem engineers}, capable of modifying 
the flows of energy and matter through the ecosystem [54,55]. This is actually an approach to restoration 
ecology that is based in the existence of multiple alternative states in complex ecosystems [56,57]. 

A major objection to developing this framework in the real natural habitats is the 
potential for evolving undesirable (or unexpected) traits. This could be labelled as the 
"Jurassic Park Effect": even designed systems aimed to population control can eventually escape 
from genetic firewalls [58]. This is a claim that is supported by the unescapable potential 
of microbial systems for evolution. However, two important points need to be made. One is that 
microbes are being constantly dispersed on a global scale without special impact on extant 
ecosystems. As it occurs with most invaders, they either fail to survive or simply become 
part of the receptor habitats, where they are over competed by resident species. 
Secondly, the design principles proposed in this paper consider engineering 
extant organisms under a cooperation-based framework (thus enhancing mutualistic loops) or 
taking advantage of human-generated waste that can act as an artificial substrate to support the 
synthetic organisms. In all cases, a synergetic interaction between design and niche context is at work. 
 
Redesigning our ecosystems requires a modification of nature, and deal 
with ecosystem complexity face to face [59].  But we should not forget 
that most biomes in our planet have already been deeply transformed by human activities [60].  
Far from what we could expect, they can be diverse, robust and more efficient in terms of 
nutrient cycling and other components of ecosystem services [61]. Despite the long, 
sustained and profound anthropogenic impact on many of these novel ecosystems, they 
can display a richness and resilience that reminds us the potential of nature to reconstruct itself. 
It is time to decide what we want and why is our role in the future of nature. If we want humans to 
be part of the biosphere, we need to foresee the future impact of climate change on our 
planet. Here too slow response can trigger shifts. In this case, social collapse [62]. 
Synthetic biology can play a major role, along with all other strategies, to modify ongoing 
trends. That means redesign nature, but perhaps too to safely exit the Anthropocene with a renewed 
relationship with ecological systems.

\vspace{0.2 cm}

{\bf Authors contributions}

RS conceived the original idea. All authors contributed to develop it, made a literature research and 
designed the final figures. RS, in close collaboration with SDN and RM, wrote the manuscript, which 
was approved by all authors . 

\vspace{0.2 cm}

\vspace{0.2 cm}

{\bf Acknowledgments}

\vspace{0.2 cm}
The authors thank the members of the Complex Systems Lab for useful discussions. 
This study was supported by an European Research Council Advanced Grant, 
the Botin Foundation, by Banco Santander through its Santander Universities Global Division 
and by the Santa Fe Institute, where most of the work was done.

\vspace{0.2 cm}

\section{References}

\begin{enumerate}

\item
 Barnovsky AD et al (2012) Approaching a state of shift in Earths biosphere. Nature 486: 52-58.
 
\item
Hughes TP et al (2013) Multiscale regime shifts and planetary boundaries. Trends Ecol. Evol. 28: 389-395.
   
\item
Scheffer M, Carpenter S, Foley JA et al (2001)  Catastrophic shifts in ecosystems. Nature 413: 591-596.
   
\item
Scheffer M (2009) Critical transitions in nature and society (Princeton U. Press, Princeton)

\item
Lenton, T.M. et al. (2008) Tipping elements in the Earths climate system. Proc. Natl. Acad. Sci. U.S.A. 105: 1786-1793

\item
Edenhofer O, Pichs?Madruga R, Sokona Y et al, editors (2011) IPCC Special Report on Renewable Energy Sources and Climate Change Mitigation. 
Cambridge University Press, Cambridge,  New York. 
  
\item
Schneider SH (2008) Geoengineering: could we or should we make it work? Phil. Trans. R. Soc. A 366: 3843-3862.
  
\item
Vaughan NE and Lenton TM (2011) A review of climate geoengineering proposals. Climatic change 109: 745-790.

\item
Caldeira K, Bala G and Cao L (2013) The science of geoengineering. Annu. Rev. Earth Planet. Sci. 41: 231-256.

\item
Glavovic BC and Smith GP (2014) {\em Adaptating to climate change}. Springer,  Dortrech.

\item
Keith DW (2000) Geoengineering the climate: history and prospect. Annu Rev Energy Environ. 25: 245-284.

\item
Rogeij J, McCollum DL, Reisinger A et al (2014) 
Probabilistic cost estimates for climate change mitigation. Nature 493: 79-83.

\item
Sol\'e R (2015) Bioengineering the biosphere? Ecol. Complexity, in press.

\item
Khalil AS and Collins JJ (2010) Synthetic biology: applications come of age. Nature Rev Genetics 11: 367-379.

\item
Weber W and Fussenegger M (2012) Emerging biomedical applications of synthetic biology. 
Nature rev Genet 13: 21-35. 

\item
Renda BA, Hammerling MJ and Barrick JE (2014) Engineering reduced evolutionary potential for synthetic biology. Mol. Biosys. 10: 1668-1878.

\item
Mandell DJ, Lajoie MJ, Mee MT et al (2015) Biocontainment of genetically modified organisms by synthetic 
protein design. Nature 518: 55-60.

\item
Simberloff D and Rejm\'anek M (eds.) (2011) Encyclopedia of biological invasions. University of california Press. Berkeley CA.

\item
Sax DF, Stachowicz JJ, Brown JH  et al (2007) Ecological and evolutionary insights from species invasions. Trends Ecol Evol 22: 465-471.

\item
Shou W, Ram S and Vilar JMG (2007) Synthetic cooperation in engineered yeast populations. Proc Natl Acad Sci USA 104: 1877-1882.

\item
Marchetti M, Cappella D, Glew M et al. (2010) Experimental evolution of a plant pathogen into a legume symbiont. 
PLoS Biology 8: e1000280.

\item
Guan SH, Gris C, Cruveiller A et al (2013)  Experimental evolution of nodule intracellular infection in legume symbionts. ISME J 7: 1367-1377.

\item 
Kiers ET, Duhamet M, Beesetty Y et al (2011) Reciprocal rewards stabilize cooperation in the mycrorrizal symbiosis.  Science 333: 880-882.

\item
Hom  EFY and Murray AW (2014) Niche engineering demonstrates a latent capacity for fungal-algal mutualism. 
Science 345: 94-98.

\item
Hillesland KL, Sujun L, Flowers JJ et al. (2014) Erosion of functional independence early in the evolution of 
a microbial mutualism. Proc Natl Acad Sci USA 111: 14822-12827.

\item
Pointing SB and Belnap J (2012) Microbial colonization and controls in dryland ecosystems. Nature Rev Microbiol 10: 551-562. 

\item
Mager DM and Thomas AD (2011)  Extracellular polysaccharides from cyanobacterial soil crusts: A review of their
role in dryland soil processes. J Arid Env 75: 91-97.

\item
Park C, Li X, Liang Jia R, Hur J-S: Effects of superabsorbent polymer
on cyanobacterial biological soil crust formation in laboratory. Arid Land Research and Management 2015, 29: 5571.

 \item
Belnap J and Lange OL (eds) (2003) {\em Biological soil crusts:  Structure, function and management}. Springer, Berlin. 

 \item
 Belnap J (2003) The world at your feet: desert biological soil crusts. Front Ecol Environ 1: 181?189.

 \item
Bowker  MA (2007) Biological Soil Crust Rehabilitation in Theory and Practice: An Underexploited Opportunity. Restoration Ecology 15: 13-23.

 \item
Bowker  MA, Mau RL, Maestre FT et al (2011) Functional profiles reveal unique ecological roles of various biological soil crust organisms. Functional Ecology 25: 787-795.
  
 \item
Brooker RW, Maestre FT, Callaway RM et al (2008) Facilitation in plant communities: the past, the present, and the future. J. Ecol. 96: 18-34
  
\item
K\'efi, S. et al. (2007) Spatial vegetation patterns and imminent desertification in Mediterranean arid ecosystems. Nature 449: 213-217.

\item
Scanlon, T. M., Caylor, K. K., Levin, S. A. and Rodriguez-Iturbe, I. (2007) Positive feedbacks promote power-law clustering of Kalahari vegetation. Nature 449: 209-212.

\item
Sol\'e R (2007) Scaling laws in the drier. Nature 449: 151-153.

\item
Rietkerk M and Van de Koppe J (2008) Regular pattern formation in real ecosystems. Trends Ecol Evol 23: 169-175.

\item
Poulter B, et al (2014) Contribution of semi-arid ecosystems to interannual variability of the global carbon cycle. Nature 509: 600?603. 
 
\item
Cases I and de Lorenzo V (2005) Genetically modified organisms for the environment: stories of success and failure and what we have learnt from them. Intl. Microb. 8: 213-222.
 
\item
de Lorenzo V (2008) Systems biology approaches to bioremediation. Curr Opin. Biotechnol. 19: 579-589.
 
\item
Gregory MR (2009) Environmental implications of plastic debris in marine setting. Phil. Trans. R Soc London B 364: 2013-2025.
 
\item
Barnes DKA (2002) Invasions by marine life on plastic debris. Nature 416: 808-809.

\item
Zettler ER, Mincer TJ and Amaral-Zettler LA (2013) Life in the Plastisphere: Microbial communities on plastic marine debris. Env. Sci. Tech. 47: 7127-7146.

\item
Kara Lavender Law K, Mor\'et-Ferguson S, Maximenko NA et al (2010) Plastic accumulation in the North Atlantic subtropical gyre. 
Science 329: 1185-1188.

\item
BacillaFilla: Fixing Cracks in Concrete. http://2010.igem.org/Team:Newcastle

\item
Jonkers HM et al (2010) Application of bacteria as self-healing agent for the development of sustainable concrete. 
Ecol Eng 36: 230-235.

\item
Rao MVS, Reddy VS, Hafsa Met al (2013) Bioengineered Concrete - A Sustainable Self-Healing Construction Material. 
Res J Eng Sci 2: 45-51.

\item
Newton RJ, McLellan SL, Dila DK et al (2015) Sewage Reflects the Microbiomes of Human Populations. mBio 6: e02574-14.

\item
Afshinnekoo E, Meydan C, Chowdhury S et al (2015) Geospatial Resolution of Human and Bacterial 
Diversity with City-Scale Metagenomics. Cell Syst 1: 1-15.

\item
Cho I and Blaser MJ (2012) The human micro biome: at the interface of health and disease. Nature  Rev Genet 13: 260-270.

\item
Costello EK (2012) The application of ecological theory toward an understanding of the human microbiome. Science 336: 1255-1262.

\item
Pepper JW and Rosenfeld S (2012) The emerging medical ecology of the human gut micro biome. Trends Ecol Evol 27: 381-384.

\item
Huttenhower C et al (2012) Structure, function and diversity of the healthy human microbiome. Nature 486: 207-214.
 
\item
Jones CG, Lawton JCG and Shachak M (1994) Organisms as ecosystem engineers. Oikos 69: 373-386.

\item
Jones CG, Lawton JCG and Shachak M (1997) Positive and negate effects of organisms as physical ecosystem engineers. Ecology 78 : 1946-1957.
 
\item
Seastedt TR, Hobbs RJ and Suding KN (2008) Management of novel ecosystems: are novel approaches required? Front Ecol Environ 6: 547-553.

\item
Suding KN, Gross KL and Houseman GR (2004) Alternative states and positive feedbacks in restoration ecology. Trends Ecol Evol 19: 46-53. 

\item
Crichton M (2012) {\em Jurassic Park}.  Ballantine Books, New York.

\item
Levin SA (2002) The biosphere as a complex adaptive system. Ecosystems 1: 431-436.

\item
Ellis EC, Kaplan JO, Fuller DQ et al (2013) Used planet: A global history. Proc Natl Acad Sci USA 110: 7978-7985.

\item
Marris E (2011) {\em Rambunctious garden. Saving nature in a post-wild world}  (Bloombsbury, New York). 
 
\item
Scheffer M, Westley F and Brock W. (2003) Slow response of societies to new problems: 
causes and costs. Ecosystems 6: 493-502.

\end{enumerate}

\end{document}